\documentclass[a4paper,11pt]{article}
\usepackage{pos}
\usepackage{comment}
\usepackage{lineno}
\usepackage{amsmath}
\usepackage{mathtools}
\usepackage{csquotes}
\usepackage{subcaption}
\usepackage[belowskip=0pt,aboveskip=0pt]{caption}
\newcommand\ddfrac[2]{\frac{\displaystyle #1}{\displaystyle #2}}

\let\OLDthebibliography\thebibliography
\renewcommand\thebibliography[1]{
  \OLDthebibliography{#1}
  \setlength{\parskip}{0pt}
  \setlength{\itemsep}{0pt plus 0.3ex}
}

\title{On the muon scale of air showers and its application to the AGASA data}
\ShortTitle{Muon scale and its application to AGASA data}

\author*[a,b]{Flavia Gesualdi}
\author[c]{Hans Dembinski}
\author[d]{Kenji Shinozaki}
\author[a]{Daniel Supanitsky}
\author[b]{Tanguy Pierog}
\author[e]{Lorenzo Cazon}
\author[f]{Dennis Soldin}
\author[e]{Ruben Concei\c{c}\~ao}
\author{for the Working group on Hadronic Interactions and Shower Physics (WHISP)}

\affiliation[a]{Instituto de Tecnologías en Detección y Astropartículas (CNEA, CONICET, UNSAM), San Martín, Argentina}
\affiliation[b]{Karlsruhe Institute of Technology, Institute for Astroparticle Physics, Karlsruhe, Germany}
\affiliation[c]{TU Dortmund University, Dortmund, Germany}
\affiliation[d]{National Centre for Nuclear Research, Lodz, Poland}
\affiliation[e]{Laboratório de Instrumentação e Física Experimental de Partículas - LIP and Instituto Superior Técnico - IST, Universidade de Lisboa - UL, Lisbon, Portugal}
\affiliation[f]{Bartol Research Institute, Dept. of Physics and Astronomy, University of Delaware, Delaware, USA}
\emailAdd{flavia.gesualdi@iteda.cnea.gov.ar}

\abstract{Recently, several experiments reported a muon deficit in air shower simulations with respect to the data. This problem can be studied using an estimator that quantifies the relative muon content of the data with respect to those of proton and iron Monte Carlo air shower simulations. We analyze two estimators. The first one, based on the logarithm of the mean of the muon content, is built from experimental considerations. It is ideal for comparing results from different experiments as it is independent of the detector resolution. The second estimator is based on the mean of the logarithm of the muon content, which implies that it depends on shower-to-shower fluctuations. It is linked to the mean-logarithmic mass $\left \langle \ln A \right \rangle$ through the Heitler-Matthews model. We study the properties of the estimators and their biases considering the knowns and unknowns of typical experiments. 
Furthermore, we study these effects in measurements of the muon density at $1000\,$m from the shower axis obtained by the Akeno Giant Air Shower Array (AGASA). 
Finally, we report the estimates of the relative muon content of the AGASA data, which support a muon deficit in simulations. These estimates constitute valuable additional information of the muon content of air showers at the highest energies.}

\FullConference{37$^{\rm{th}}$ International Cosmic Ray Conference (ICRC 2021)\\
		July 12th -- 23rd, 2021\\
		Online -- Berlin, Germany}

\begin{document}
\maketitle
\section{Introduction}
Cosmic rays with energies above $10^{15}\,$eV can only be efficiently studied through their extensive air showers. Such primary energies are inaccessible to the Large Hadron Collider. Therefore, high-energy hadronic interaction models can only be tested in the ultra-high energy range through measurements of air showers. Models are usually tested by analyzing the consistency of the predictions of mean logarithmic mass $\langle \ln A \rangle$ derived from two observables: the depth of the shower maximum $X_{\text{max}}$ and the muon content $N_{\mu}$. By studying this, such models can be improved, which in exchange can improve the precision of the inferred $\langle \ln A \rangle$~\cite{Auger14,Auger21}.

Different experiments reported a muon deficit in air shower simulations with respect to data at different energies. The Working group on Hadronic Interactions and Shower Physics (WHISP) analyzed this problem, regarded as \enquote{the muon puzzle}, by combining the measurements of leading air shower experiments. Data from HiRes-MIA, NEVOD-DECOR, SUGAR array, AMIGA (the muon detectors from the Pierre Auger Observatory), the Pierre Auger Observatory, and Telescope Array, are consistent with said muon deficit. On the other hand, data from EAS-MSU, KASCADE-Grande, the IceCube Neutrino Observatory, and Yakutsk array are consistent with no muon deficit. The combined analysis based on post-LHC hadronic-interaction models finds a muon deficit increasing with energy, with a non-vanishing slope at $8\,\sigma$ significance~\cite{Dembinski19,Cazon19}. This analysis did not yet include the measurements of the Akeno Giant Air Shower Array (AGASA). These can add a valuable piece to the puzzle, especially considering the high energies that these data reach. A recent re-analysis finds AGASA data consistent with a muon deficit in simulations \cite{Gesualdi20}.

The AGASA experiment was comprised of an array of 111 scintillation counters spread across $\sim 100\,\text{km}^2$. It was able to detect air showers with energies above $\sim 3\times 10^{18}\,\text{eV}$ and with zenith angles $\theta \leq 45^{\circ}$. AGASA also had 27 muon detectors which covered $\sim 30\,\text{km}^2$. These consisted of proportional counters shielded with $30\,\text{cm}$ of iron or $1\,\text{m}$ of concrete; the vertical muon energy threshold was $0.5\,\text{GeV}$~\cite{Hayashida95}. The detectors were decommissioned in 2004.  

In this work we analyze the biases of two estimators of the relative muon content of data with respect to simulations, and estimate their values from AGASA measurements.  

\section{Definition of the muon scale and muon deficit scale} \label{sec:defs}

The Heitler-Matthews model predicts that the muon content $N_{\mu}$ of an air shower grows with the mass $A$ and with the primary energy $E$ according to
\begin{equation}
\ln N_{\mu} = (1-\beta) \ln A + \beta \ln(E/\xi_{c}),
\label{eq:HM}
\end{equation}
where $\beta$ is the power-law index ($\beta \approx 0.9$) and $\xi_{c}$ is a critical energy constant~\cite{Matthews05}. 

We want to define a muon scale independent of experimental details of the measurement, such as the radial distance to shower axis, the zenith angle of the shower, the shower age, and so forth, so that measurements under different conditions can be compared. Air shower simulations approximately describe these aspects, which means that ratios of measured and simulated muon numbers (or equivalent: differences in their logarithms) are approximately independent of these experimental details. With this guiding idea in mind, the muon scale was defined in Ref.~\cite{Dembinski19} as
\begin{equation}
z \coloneqq \frac{\ln N_{\mu,\,\text{data}}^{\text{det}} - \ln N_{\mu,\,\text{p}}^{\text{det}}}{\ln N_{\mu,\,\text{Fe}}^{\text{det}} - \ln  N_{\mu,\,\text{p}}^{\text{det}}},
\label{eq:defnoavg}
\end{equation}
where the suffix \enquote{det} indicates that the proton and iron muon contents derive from full-detector simulations. This is useful to cancel first order detector effects. 
The muon scale defined in Eq.~(\ref{eq:defnoavg}) has desirable properties: In the first place, it is independent of the energy to first order (the remaning energy dependence is only through the composition). In the second place, it should range between 0 and 1 for proton-like and iron-like data respectively, given that simulations correctly reproduce the muon content of real showers. Finally, if the mass number $A$ is known event-by-event, the Heitler-Matthews model predicts a value of $z$ of $\ln A /\ln 56$. 

To build an estimator of the muon scale $z$, it is necessary to introduce the mean within a given energy bin. We define the first estimator of the muon scale by taking the logarithm of the mean of the muon number,
$z_{\ln \langle \cdot \rangle} = (\ln \langle N_{\mu,\,\text{data}}^{\text{det}} \rangle - \ln \langle N_{\mu,\,\text{p}}^{\text{det}}\rangle)/(\ln \langle N_{\mu,\,\text{Fe}}^{\text{det}}\rangle - \ln  \langle N_{\mu,\,\text{p}}^{\text{det}}\rangle)$.
The second estimator is defined by taking the mean of the logarithm of the muon number, 
$z_{\langle \ln \cdot \rangle} = (\langle \ln N_{\mu,\,\text{data}}^{\text{det}} \rangle - \langle \ln N_{\mu,\,\text{p}}^{\text{det}}\rangle)/(\langle \ln N_{\mu,\,\text{Fe}}^{\text{det}}\rangle - \langle \ln N_{\mu,\,\text{p}}^{\text{det}}\rangle)$. It is important to stress that the logarithm of the mean or the mean of the logarithm should be used consistently. Any \enquote{mixed} estimator would suffer from undesired biases.

These two estimators are in principle different, since the logarithm of the mean and the mean of the logarithm of the muon content are not the same. One can be expressed as a function of the other by using that $N_{\mu} = \langle N_{\mu} \rangle (1 + \epsilon)$, with $\epsilon  = (N_{\mu} - \langle N_{\mu} \rangle)/\langle N_{\mu} \rangle$. Then
\begin{align}
\langle \ln N_{\mu} \rangle &= \ln \langle N_{\mu} \rangle  + \left\langle\ln\left(1 + \epsilon\right) \right\rangle = \ln \langle N_{\mu} \rangle + \left \langle \epsilon - \tfrac{1}{2} \epsilon^2+ \mathcal{O} (\epsilon^3) \right\rangle,\\
&\approx \ln \langle N_{\mu} \rangle - \tfrac{1}{2} \left(\text{RSD}_{\text{tot}}[N_\mu]\right)^2,
\label{eq:lnapprox}
\end{align}
where $\left \langle \epsilon^2 \right \rangle = \left(\sigma_{\text{tot}}(N_\mu)/\left\langle N_{\mu} \right\rangle\right)^2  = \left(\text{RSD}_{\text{tot}}[N_\mu]\right)^2$ is the square of the relative standard deviation of $N_\mu$ \cite{Dembinski18}. The suffix \enquote{tot} marks that all sources of fluctuations have to be considered.

Furthermore, to evaluate the muon deficit, we need to consider a reference $z$-value. Its expected value is $z_{\text{mass}} \coloneqq (\ln N_{\mu,\,\text{mass}}^{\text{det}} -\ln N_{\mu,\,\text{p}}^{\text{det}})/(\ln N_{\mu,\,\text{Fe}}^{\text{det}} - \ln  N_{\mu,\,\text{p}}^{\text{det}})$, where $N_{\mu,\,\text{mass}}^{\text{det}}$ represents the muon content from detector simulations, evaluated using the mass fractions of the composition model. If the measured $z$-values follow $z_{\text{mass}}$, then simulations have no muon deficit. Therefore, a scale of the muon deficit in simulations, $\Delta z$, is defined as
$\Delta z \coloneqq z - z_{\text{mass}}$. 

In this work we take as a reference $z_{\langle \ln \cdot \rangle\,\text{mass}}^{\text{HM}} = \langle \ln A \rangle/\ln 56$. This expression can be obtained by superposition of the Heitler-Matthews model. For this, shower-to-shower fluctuations are introduced through variations in $\beta$, and a mixed composition through different values of $A$. After the superposition, $\langle \ln N_{\mu} \rangle$ keeps a linear relation to $\langle \ln A \rangle$. This is also supported by simulations \cite{Dembinski18}.

\section{Biases of the muon scale and muon deficit estimators}\label{sec:biases}
The estimators of the muon scale, $z_{\ln \langle \cdot \rangle}$ and $z_{\langle \ln \cdot \rangle}$, can be biased if the detector simulations mismodel the real detector effects in $\langle N_{\mu,\{\text{p,Fe}\}}^{\text{det}} \rangle$, or if there is a composition bias that affects $\langle N_{\mu,\text{data}}^{\text{det}} \rangle$. These impact on both estimators approximately to the same degree. These sources of bias are contemplated in the systematic uncertainties.

$z_{\ln \langle \cdot \rangle}$ is a better estimator of the muon scale defined in Eq.~(\ref{eq:defnoavg}) because it does not explicitly depend on the detector resolution. In other words, $z_{\ln \langle \cdot \rangle}$ can be compared among experiments.

In contrast, $z_{\langle \ln \cdot \rangle}$ suffers from systematics from a possibly mis-modeled (\enquote{mm}) detector resolution. To understand this effect, we assume that $\left\langle N_{\mu,\text {p}}^{\text{mm}}\right\rangle \approx \left\langle N_{\mu,\text {p}}\right\rangle$, since this source of bias is already accounted for in the systematic uncertainties. The total detector resolution depends on the muon number detector resolution $\text{RSD}_{\text{det}}[N_{\mu}]$, and on the energy resolution $\text{RSD}_{\text{det}}[E]$. Both affect the measured $N_{\mu}(E)$. We can use Eq.~(\ref{eq:HM}) to propagate the energy resolution and obtain an approximation of the total muon number resolution 
$\text{RSD}_{\text{det}\oplus}[N_{\mu}] \approx \sqrt{(\text{RSD}_{\text{det}}[N_{\mu}])^2 + (\beta \cdot \text{RSD}_{\text{det}}[E])^2},$
where $\beta \approx 0.9$. If the detector simulations mismodel the total detector resolution, the estimate of $z_{\left\langle \ln \cdot \right\rangle}$ is also mismodeled: $z_{\left\langle \ln \cdot \right\rangle}^{\text{mm}} = \frac{\left\langle\ln N_{\mu,\text{data}} \right\rangle-\left\langle\ln N_{\mu,\text {p}}^{\text{mm}} \right\rangle}{\left\langle\ln N_{\mu,\text{Fe}}^{\text{mm}} \right\rangle-\left\langle\ln N_{\mu,\text {p}}^{\text{mm}}\right\rangle}$. Using Eq.~(\ref{eq:lnapprox}) to give approximate expressions of $\left\langle\ln N_{\mu,\{\text{p,Fe}\}}^{\text{mm}}\right\rangle$ and $\left\langle\ln N_{\mu,\{\text{p,Fe}\}}\right\rangle$, and using Eq.~(\ref{eq:HM}) to approximate $\langle\ln(N_{\mu,\text{Fe}})\rangle-\langle\ln(N_{\mu,\text {p}})\rangle \approx (1-\beta) \ln(56)$, 
$z_{\left\langle \ln \cdot \right\rangle}^{\text{mm}} - z_{\left\langle \ln \cdot \right\rangle}$ can be expressed as
\begin{equation}
z_{\left\langle \ln \cdot \right\rangle}^{\text{mm}} - z_{\left\langle \ln \cdot \right\rangle} \approx  \frac{\text{RSD}_{\text{det}\oplus}[N_{\mu}] \left( \text{RSD}_{\text{det}\oplus}^{\text{mm}}[N_{\mu}] - \text{RSD}_{\text{det}\oplus}[N_{\mu}]\right) 
+\tfrac{1}{2}
 \left(\text{RSD}_{\text{det}\oplus}^{\text{mm}}[N_{\mu}]- \text{RSD}_{\text{det}\oplus}[N_{\mu}]\right) ^2 }{(1-\beta)\ln(56)}.
\end{equation}

\begin{figure}[h!]
\centering
\includegraphics[width=0.75\textwidth]{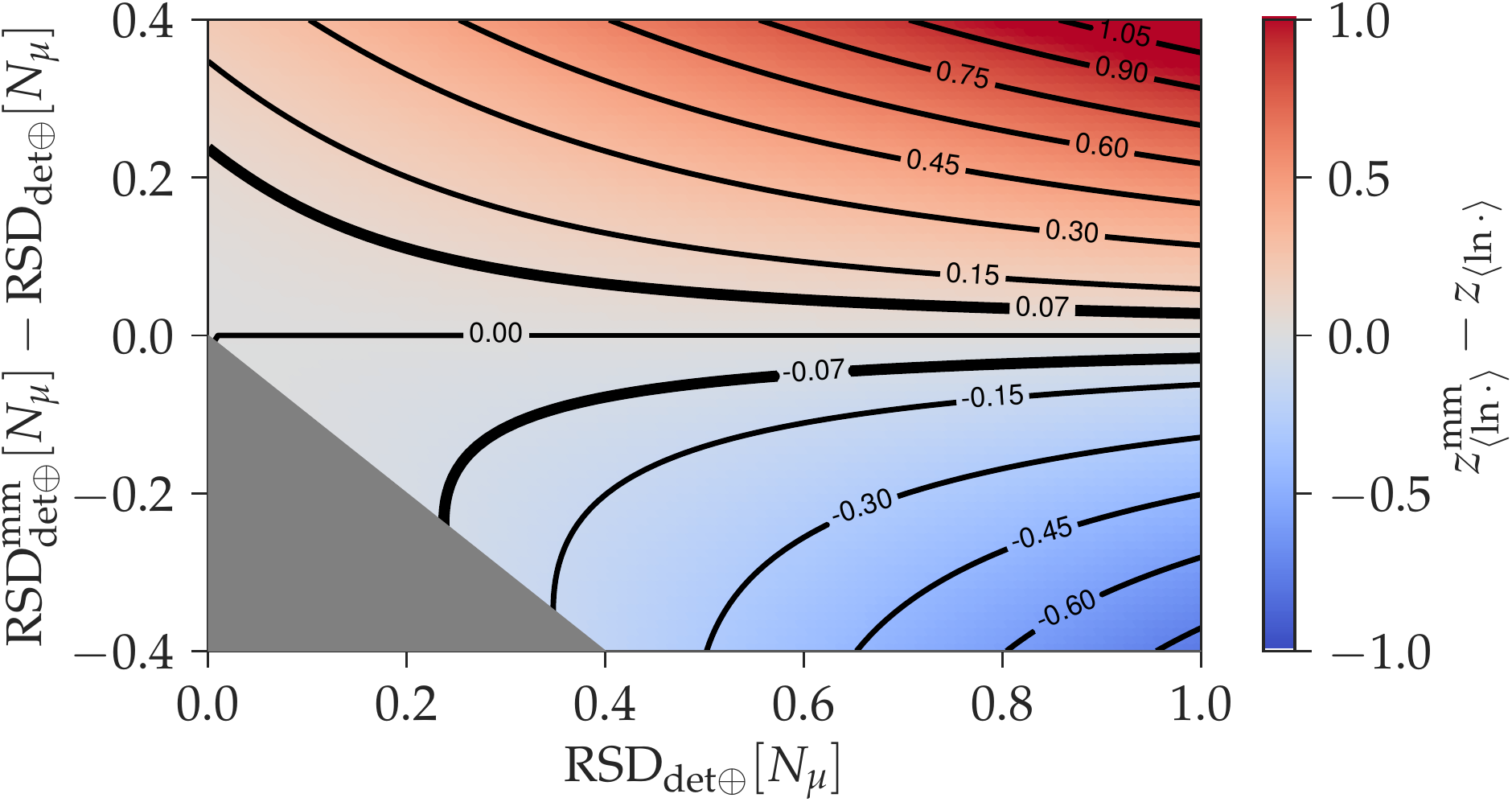}
\caption{$z_{\left\langle \ln \cdot \right\rangle}^{\text{mm}} - z_{\left\langle \ln \cdot \right\rangle}$ (color scale and contours) as a function of the true total detector resolution (x-axis), and of the difference between the mismodeled and true total detector resolutions
(y-axis). The region where $\text{RSD}_{\text{det}\oplus}^{\text{mm}}[N_{\mu}] \leq 0$ is shaded in gray. The thick countours correspond to $|z_{\left\langle \ln \cdot \right\rangle}^{\text{mm}} - z_{\left\langle \ln \cdot \right\rangle}| = 0.07$.} 
\label{fig:detres}
\end{figure}

In Fig.~\ref{fig:detres} we plot $z_{\left\langle \ln \cdot \right\rangle}^{\text{mm}} - z_{\left\langle \ln \cdot \right\rangle}$ (color scale) as a function of the true total detector resolution $\text{RSD}_{\text{det}\oplus}[N_{\mu}]$ (x-axis) and the difference between the mismodeled and true total detector resolutions $\text{RSD}_{\text{det}\oplus}^{\text{mm}}[N_{\mu}] - \text{RSD}_{\text{det}\oplus}[N_{\mu}]$ (y-axis). $z_{\left\langle \ln \cdot \right\rangle}^{\text{mm}} - z_{\left\langle \ln \cdot \right\rangle}$ equals zero when the modelled resolution equals the true one, as expected. A gray triangle shades the unphysical region where $\text{RSD}_{\text{det}\oplus}^{\text{mm}}[N_{\mu}] \leq 0$. The thick contours correspond to a systematic in $z_{\left\langle \ln \cdot \right\rangle}$ of $\pm 0.07$. 
For example, for a true detector resolution of $30\,\%$, a systematic of $\pm 0.07$ in $z_{\left\langle \ln \cdot \right\rangle}$ is attained when the modelled resolution is $8\,\%$ larger or $12\,\%$ smaller than the true one. Ref.~\cite{Auger21} is a recent example in which the mismodelling of detector fluctuations was investigated.
\vspace{-0.5em}

Having analyzed the estimators of the muon scale $z$, we continue to assess the quality of $\Delta z_{\ln \langle \cdot \rangle}$ and $\Delta z_{\langle \ln \cdot \rangle}$ as estimators of the muon deficit, when using $z_{\langle \ln \cdot \rangle\,\text{mass}}^{\text{HM}}$ as a reference. 

The systematic error in $\Delta z_{\langle \ln \cdot \rangle}$ propagates from $z_{\langle \ln \cdot \rangle}$, which is already understood, and from $z_{\langle \ln \cdot \rangle\,\text{mass}}^{\text{HM}}$. The latter suffers from a systematic error propagated from that of the composition (like any estimator of $z_{\text{mass}}$). It also has an additional (smaller) systematic error due to the Heitler-Matthews model not reproducing simulations perfectly. 

In the case of $\Delta z_{\ln \langle \cdot \rangle}$, $z_{\langle \ln \cdot \rangle\,\text{mass}}^{\text{HM}}$ introduces an additional systematic error, because it is the predicted value of $z_{\langle \ln \cdot \rangle\,\text{mass}}=(\langle \ln N_{\mu,\,\text{mass}}^{\text{det}} \rangle - \langle \ln N_{\mu,\,\text{p}}^{\text{det}}\rangle)/(\langle \ln N_{\mu,\,\text{Fe}}^{\text{det}}\rangle - \langle \ln N_{\mu,\,\text{p}}^{\text{det}}\rangle)$
instead of $z_{\ln \langle \cdot \rangle\,\text{mass}} = (\ln \langle N_{\mu,\,\text{mass}}^{\text{det}} \rangle - \ln \langle N_{\mu,\,\text{p}}^{\text{det}}\rangle)/(\ln \langle N_{\mu,\,\text{Fe}}^{\text{det}}\rangle - \ln  \langle N_{\mu,\,\text{p}}^{\text{det}}\rangle)$. The difference between these two can be understood by using Eq.~(\ref{eq:lnapprox}) to approximate each term in $z_{\langle \ln \cdot \rangle\,\text{mass}}$ as 
\begin{equation}
z_{\langle \ln \cdot \rangle\,\text{mass}} \approx \frac{\ln \langle N_{\mu,\,\text{mass}}^{\text{det}} \rangle - \ln \langle N_{\mu,\,\text{p}}^{\text{det}}\rangle - \tfrac{1}{2} \left[\left( \text{RSD}_{\text{sh-sh}}[N_{\mu,\,\text{mass}}] \right)^{2} - \left( \text{RSD}_{\text{sh-sh}}[N_{\mu,\,\text{p}}] \right)^{2}\right]}{\ln \langle N_{\mu,\,\text{Fe}}^{\text{det}}\rangle - \ln  \langle N_{\mu,\,\text{p}}^{\text{det}}\rangle - \tfrac{1}{2} \left[\left( \text{RSD}_{\text{sh-sh}}[N_{\mu,\,\text{Fe}}] \right)^{2} - \left( \text{RSD}_{\text{sh-sh}}[N_{\mu,\,\text{p}}]\right)^{2}\right]},
\label{eq:approxzlnamasstozmass}
\end{equation}
where we used that only shower-to-shower fluctuations depend on the primary, and all other sources of fluctuations cancel out. 
Comparing $z_{\ln \langle \cdot \rangle\,\text{mass}}$ to Eq.~(\ref{eq:approxzlnamasstozmass}) we see that $z_{\langle \ln \cdot \rangle\,\text{mass}}$ depends on how the muon content fluctuates shower-to-shower, while $z_{\ln \langle \cdot \rangle\,\text{mass}}$ does not. These fluctuations affect $z_{\langle \ln \cdot \rangle\,\text{mass}}^{\text{HM}}$, and therefore $\Delta z_{\ln \langle \cdot \rangle}$, introducing a bias. This source of bias in $\Delta z_{\ln \langle \cdot \rangle}$ is largest when $\left(\text{RSD}_{\text{sh-sh}}[N_{\mu,\,\text{mass}}]\right)^{2}$ is maximum, which happens approximately at a $50\,\%$ proton - $50\,\%$ iron mixture. 
It was reported to be of 0.07 in the worst case scenario~\cite{Dembinski18}. This bias is in general smaller than the possible systematic errors in $\Delta z_{\langle \ln \cdot \rangle}$, considering the unknowns of typical experiments. Furthermore, this bias can be corrected if shower-to-shower flucutations for single primaries are known (a parameterization can be found in Ref.~\cite{Dembinski18}), as well as $\sigma(\ln A)$. The latter are predicted by the Global Spline Fit (GSF) \cite{Dembinski17} and the Pierre Auger \cite{Bellido17} composition models. 
\vspace{-0.5em}

\section{The muon scale from AGASA data}
The muon density at $1000\,\text{m}$ from the shower axis, measured on the shower plane, in air showers detected by AGASA are extracted from Fig.~7 of Ref.~\cite{Shinozaki04} (see table in appendix B of Ref.~\cite{Gesualdi20}), and are grouped into three reconstructed energy bins of width $\Delta \log_{10} (E_R/\text{eV}) = 0.2$. The energy associated to each muon density measurement is brought to the reference energy scale defined by the \textit{Spectrum Working Group}~\cite{Ivanov17} by multiplying it by a factor of $0.68$~\cite{Gesualdi20}. Most of the difference between the energy scales originates from the muon deficit in the simulations used to calibrate the AGASA energy scale (especially the older-generation ones). The selected events have rescaled reconstructed energies in $18.83\,\leq\,\log_{10}(E_{R}/\textrm{eV})\,\leq\,19.46$) and zenith angles $\theta\leq36^{\circ}$.

Furthermore, we use a library of proton, helium, nitrogen, and iron initiated air showers, simulated using the models QGSJetII-04, EPOS-LHC, and Sibyll2.3c with CORSIKA. This library is described in Ref.~\cite{Gesualdi20}. For every primary and interaction model, the mean muon density as a function of the energy is fitted with a power-law. We mimic the effects of the energy reconstruction and energy binning analytically. The mean simulated muon density, calculated in the $i$-th reconstructed energy bin $E_{Ri}$, can be expressed as\\
\vspace{-0.75em}
\begin{equation}
\left\langle \rho_\mu (E_{Ri})\right\rangle = \ddfrac{\int_{E_{Ri}^{-}}^{E_{Ri}^{+}}dE_{R} \int_{0}^{\infty}dE\ 
\langle \widetilde{\rho}_\mu(E)\rangle \cdot J(E)\cdot G(E_{R}|E) }{\int_{E_{Ri}^{-}}^{E_{Ri}^{+}} dE_{R}
\int_{0}^{\infty}dE\  J(E)\cdot G(E_{R}|E)},
\label{eq:conv}
\end{equation}\\
where $E_{Ri}$ is the center the reconstructed energy bin, $E_{Ri}^{-}$ and $E_{Ri}^{+}$ are the lower and upper limits of that bin. Here, $\langle \widetilde{\rho}_\mu(E)\rangle$ is the mean muon density as a function of the true or input energy of the 
simulation (the power-law fits); $J(E)$ is the cosmic ray flux, which is obtained from a fit to the Pierre Auger Observatory measurements \cite{Fenu17}; and $G (E_{R}|E)$ is the conditional probability distribution of $E_R$ conditioned to $E$, which is reported to be a log-normal distribution \cite{Takeda03} with a standard deviation that decreases with energy \cite{Yoshida95}. A more extensive explanation of the rationale behind Eq.(\ref{eq:conv}) can be found in Ref.~\cite{Gesualdi20}. Evaluated at a specific numerical value $E^{*}$, $\left\langle \rho_\mu (E_{Ri}=E^{*})\right\rangle$ can be $8\,\%$ to $15\,\%$ smaller than $\langle \widetilde{\rho}_\mu(E = E^{*})\rangle$ within the analyzed energy range.

Having the muon density from data and from the simulations through Eq.~(\ref{eq:conv}), we can estimate $z_{\ln \langle \cdot \rangle}$ directly as defined in Sec.~\ref{sec:defs}; Note that all equations of Secs.~(\ref{sec:defs},\ref{sec:biases}) hold true when replacing $N_{\mu}$ with $\rho_{\mu}$. In contrast, the computation of $z_{\langle \ln \cdot \rangle}$ cannot be done directly as defined in Sec.~\ref{sec:defs}. This is because the values of $\left\langle \ln [\rho_{\mu\,\{\text{p,Fe}\}}(E_{Ri})]\right\rangle$ cannot be analytically computed (like in Eq.~(\ref{eq:conv})) without a very detailed knowledge of the detector resolution.

However, we can use Eq.~(\ref{eq:lnapprox}) to give an estimate of $z_{\langle \ln \cdot \rangle}$. The idea is to separate the sources of fluctuations into primary-dependent (shower-to-shower fluctuations), and primary-independent (all other sources approximately). Therefore, using the approximation from Eq.~(\ref{eq:lnapprox}) for $\left\langle \ln \rho_{\mu\,\{\text{p,Fe}\}}\right\rangle$ in the definition of $z_{\langle \ln \cdot \rangle}$ we obtain
\begin{equation}
z_{\langle \ln \cdot \rangle} \approx 
\frac{
\left\langle \ln \rho_{\mu,\text{data}}^{\text{det}} \right\rangle-\ln\left\langle \rho_{\mu,\text{p}}\right\rangle + \tfrac{1}{2}\left[\left(
\text{RSD}_{\text{sh-sh}}[\rho_{\mu,\,\text{p}}]\right)^2 + \left(\text{RSD}_{\text{not sh-sh}}[\rho_{\mu}]\right)^2\right]}
{\ln\left\langle \rho_{\mu,\text{Fe}} \right\rangle-\ln\left\langle \rho_{\mu,\text{p}}\right\rangle + \tfrac{1}{2}\left[\left(\text{RSD}_{\text{sh-sh}}[\rho_{\mu,\,\text{p}}]\right)^2 - \left(\text{RSD}_{\text{sh-sh}}[\rho_{\mu,\,\text{Fe}}]\right)^2\right]},
\label{eq:zlnaagasa}
\end{equation}
where we use in the denominator that all sources of fluctuations except shower-to-shower cancel out. In the numerator, we also split the total relative variance of proton air showers into shower-to-shower and other fluctuations. The second is computed as $\left(\text{RSD}_{\text{not sh-sh}}[\rho_{\mu}]\right)^2 = \left(\text{RSD}_{\text{tot}}[\rho_{\mu,\,\text{data}}]\right)^2 - \left(\text{RSD}_{\text{sh-sh}}[\rho_{\mu,\,\text{mass}}]\right)^2,$
where $\text{RSD}_{\text{tot}}[\rho_{\mu,\,\text{data}}]$ is computed from the data scatter within each reconstructed energy bin, while $\text{RSD}_{\text{sh-sh}}[\rho_{\mu,\,\text{mass}}]$ is computed from simulations assuming a mixed composition. The latter is obtained by fitting a normal distribution to the weighted sum of the single-primary distributions of the simulated muon densities, where the weights are the mass fractions given by the GSF model. The dependence of this approximation of $z_{\langle \ln \cdot \rangle}$ with a composition model is an evident disadvantage and source of uncertainties.

Statistical uncertainties are propagated into both estimates of $z$ from those of the muon density of data and simulations. Systematic uncertainties have several sources. The choice of a specific flux parameterization in Eq.~(\ref{eq:conv}) introduces a systematic uncertainty of $\sim 1\,\%$ in the muon density itself \cite{Gesualdi20}, which is negligible against other sources. The systematics in the reconstructed energy are computed as the sum in quadrature of three contributions. The first one amounts to a $\pm 7\,\%$ systematic in energy \cite{Yoshida94}, and arise from a possible bias in the lateral distribution of muons (from the used empirical function, the exclusion of zero-density data, and the absence of non-hit detectors). The second contribution, which amounts for a $17\,\%$ to $20\,\%$  systematic in energy (estimated from Fig. 17 of Ref.~\cite{Yoshida94}), arises from the constant intensity cut method in the zenith angle range of this data set ($0^{\circ} \leq \theta \leq 36^{\circ}$). The third contribution, ranging between $8\,\%$ and $10\,\%$, comes from the exponent in the energy scale formula. This exponent is taken from simulations, and different interaction models predict different values for the exponent \cite{Takeda03}. Finally, there is also a $\sim\!10\,\%$ uncertainty associated with the reference energy scale, which we treat separately as it is the same for any experiment in this scale. The energy systematics are added in quadrature with a factor $\beta=0.9$ (see Sec.~\ref{sec:biases}) to the intrinsic systematics of the muon density, to obtain the total systematics of the muon density ($18\,\%$ to $21\,\%$). Then these are propagated into each estimate of $z$. 

\begin{figure}[htbp]
  \includegraphics[width=\textwidth]{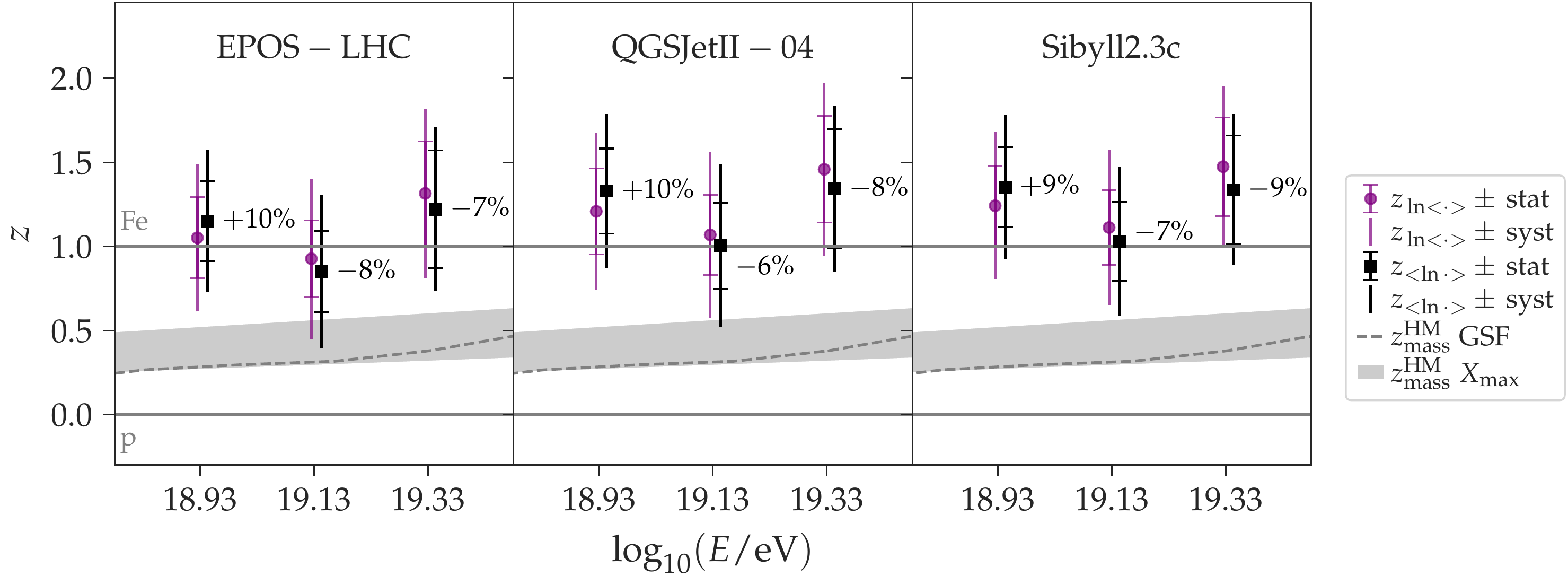}
\caption{$z_{\ln \langle \cdot \rangle}$ (circles) and $z_{\langle \ln \cdot \rangle}$ (squares), for EPOS-LHC (left), QGSJetII-04 (center), and Sibyll2.3c (right) as a function of the logarithmic energy. A horizontal displacement between the two sets is introduced for clarity. The labels show the relative difference of $z_{\langle \ln \cdot \rangle}$ to $z_{\ln \langle \cdot \rangle}$. We also show $z_{\text{mass}}^{\text{HM}}$ from the GSF model (dashed line) and from the Pierre Auger $X_{\text{max}}$ (shaded area).}
\label{fig:zvszlnaagasa}
\end{figure}

In Fig.~\ref{fig:zvszlnaagasa} we show the comparison between the estimates of $z_{\ln \langle \cdot \rangle}$ (as defined in Sec.~\ref{sec:defs}) and $z_{\langle \ln \cdot \rangle}$ (as in Eq.~(\ref{eq:zlnaagasa})). The relative difference of $z_{\langle \ln \cdot \rangle}$ to $z_{\ln\langle \cdot \rangle}$ (see labels) ranges from $-9\,\%$ to $+10\,\%$. The values of $z_{\text{mass}}^{\text{HM}}$ from the GSF and Pierre Auger $X_{\text{max}}$ composition models are also shown. By substracting the first one from $z_{\ln \langle \cdot \rangle}$ and $z_{\langle \ln \cdot \rangle}$, we obtain the values of $\Delta z_{\ln\langle \cdot \rangle}$ and $\Delta z_{\langle \ln \cdot \rangle}$.

\vspace{-0.5em}

\vspace{-0.5em}

The difference between the estimators can be understood from Sec.~(\ref{sec:biases}). We know $z_{\langle \ln \cdot \rangle}$ suffers from a systematic error if the detector resolution is mismodeled. The estimated total detector resolution is part of $\left(\text{RSD}_{\text{not sh-sh}}[\rho_{\mu}]\right)$ in Eq.~(\ref{eq:zlnaagasa}), where it is grouped with all primary-independent sources of fluctuations, such as the different energies contributing to a same bin. By normalizing muon densities to the center of the reconstructed energy bin and substracting $\left(\text{RSD}_{\text{sh-sh}}[\rho_{\mu,\,\text{mass}}]\right)^2$, we obtain an estimated total detector resolution ranging between $50\,\%$ to $84\,\%$ (depending on the energy). From Fig.~\ref{fig:detres} we can see that a total detector resolution of $50\,\%$ mismodeled in as little as $\sim\!\pm 6\,\%$ translates into a systematic in $z_{\langle \ln \cdot \rangle}$ and $\Delta z_{\langle \ln \cdot \rangle}$ of already $\pm 0.07$. Our estimate of the the total detector resolution is not more precise than $\pm 6\,\%$. In conclusion, for the AGASA data analyzed in this work, $z_{\ln \langle \cdot \rangle}$ and $\Delta z_{\ln \langle \cdot \rangle}$ are better estimators of the muon scale and muon deficit, compared to $z_{\langle \ln \cdot \rangle}$ and $\Delta z_{ \langle \ln \cdot \rangle}$. 

The $z_{\ln \langle \cdot \rangle}$ and $\Delta z_{\ln \langle \cdot \rangle}$ values computed from AGASA data are compared to those of other experiments in Ref.~\cite{Soldin21}. The agreement with the Pierre Auger findings is remarkable. The values are larger than those of Yakutsk Array but are compatible within total uncertainties. The $\Delta z_{\ln \langle \cdot \rangle}$ values from AGASA data support a muon deficit in simulations at the highest energies. 

\section{Conclusions}
The objective of this work is to study two estimators of the muon scale and compute their values from the muon density measurements at high energies from AGASA data. The first estimator, $z_{\ln \langle \cdot \rangle}$, is computed from the logarithm of the mean of the muon number/density in data and simulations. The second estimator,  $z_{\langle \ln \cdot \rangle}$, is computed from the mean of the logarithm of the muon number/density. To estimate the muon deficit, both are referenced to $z_{\text{mass}}^{\text{HM}} = \langle \ln A \rangle / \ln 56$.

For comparing the muon scale of different experiments, $z_{\ln \langle \cdot \rangle}$ is better, because its value is independent of the detector resolution. In contrast, $z_{\langle \ln \cdot \rangle}$ is subject to systematics from a mismodeled detector resolution. This systematic error depends on the total detector resolution and the degree to which it is mismodeled. It also translates directly into an error in $\Delta z_{\langle \ln \cdot \rangle}$. Moreover, $\Delta z_{\ln \langle \cdot \rangle}$ is biased from the shower-to-shower fluctuations in $z_{\text{mass}}^{\text{HM}}$, at most in $\sim 0.07$. Typically, this bias is smaller than the systematic error in $\Delta z_{\langle \ln \cdot \rangle}$, and can be corrected for if $\sigma( \ln A )$ is known. Therefore, $\Delta z_{\ln \langle \cdot \rangle}$ proves a better estimator of the muon deficit. This applies to the analyzed AGASA data. 

Finally, we compute the values of $z_{\ln \langle \cdot \rangle}$ and $\Delta z_{\ln \langle \cdot \rangle}$ from the AGASA data. We obtain that $z_{\ln \langle \cdot \rangle}\in [0.9,1.4]$ and $\Delta z_{\ln \langle \cdot \rangle} \in [0.6, 1.1] $. The AGASA data are in very good agreement with the Pierre Auger data, and are larger than the Yakutsk array values, but compatible within total uncertainties. The AGASA data support a muon deficit in simulations. These estimates add a valuable quantification of the muon scale, and constitute evidence of a muon deficit in simulations at the highest energies.

\vspace{-0.5em}
\section*{Acknowledgements}
\vspace{-0.5em}
Work by F.~Gesualdi was partially supported by the iProgress scholarship from the Helmholtz Alliance for Astroparticle Physics. This work was done in cooperation with the WHISP group, which is a joint effort by the EAS-MSU, IceCube, KASCADE-Grande, NEVOD-DECOR, Pierre Auger, SUGAR, Telescope Array and Yakutsk EAS Array collaborations.


%
%
%

\end{document}